\documentclass[aps,prc,preprint,amsmath,amssymb,showpacs,preprintnumbers,superscriptaddress]{revtex4-1}
\usepackage{CJK}
\usepackage{graphicx}% Include figure files
\usepackage{dcolumn}% Align table columns on decimal point
\usepackage{bm}% bold math
\usepackage{mathrsfs}   % added by S.B. Wang to use \mathscr{L}_{N\pi}
\usepackage{color,xcolor}  % added by S.B. Wang to use \color{red}
\usepackage{multirow}  % added by S.B. Wang to use multirow in table
\usepackage{booktabs} % added by S.B. Wang to use \cmidrule(r){5-6}
 
\usepackage{hyperref}% add hypertext capabilities
\newcommand{\bbm}{\begin{bmatrix}}
\newcommand{\ebm}{\end{bmatrix}}
\newcommand{\bBm}{\begin{Bmatrix}}
\newcommand{\eBm}{\end{Bmatrix}}
\newcommand{\bpm}{\begin{pmatrix}}
\newcommand{\epm}{\end{pmatrix}}

\begin{document}
%\begin{CJK*}{UTF8}{song} % Use default fonts from CJK (see below)


% \title{Confronting relativistic \textit{ab initio} equations of state with universal characteristics of neutron star matter}
\title{Exploring universal characteristics of neutron star matter with relativistic \textit{ab initio} equations of state}

\author{Sibo Wang}
% \email{sbwang@cqu.edu.cn}
\affiliation{Department of Physics, Chongqing University, Chongqing 401331, China}

\author{Chencan Wang}
% \email{chencanwang@mail.nankai.edu.cn}
\affiliation{School of Physics, Nankai University, Tianjin 300071, China}

\author{Hui Tong}
\email{huitong@tjnu.edu.cn}
\affiliation{College of Physics and Materials Science, Tianjin Normal University, Tianjin 300387, China}
\affiliation{Strangeness Nuclear Physics Laboratory, RIKEN Nishina Center, Wako, 351-0198, Japan}

\date{\today}

\begin{abstract}

Starting from the relativistic realistic nucleon-nucleon ($NN$) interactions, the relativistic Brueckner-Hartree-Fock (RBHF) theory in the full Dirac space is employed to study neutron star properties.
First, the one-to-one correspondence relation for gravitational redshift and mass is established and used to infer the masses of isolated neutron stars by combining gravitational redshift measurements.
Next, the ratio of the moment of inertia $I$ to mass times radius squared $MR^2$ as a function of the compactness $M/R$ is obtained, and is consistent with the universal relations in the literature.
The moment of inertia for $1.338M_\odot$ pulsar PSR J0737-3039A $I_{1.338M_\odot}$ is predicted to be 1.356$\times10^{45}$, 1.381$\times10^{45}$, and $1.407\times10^{45}\ \mathrm{g~cm^2}$ 
by the RBHF theory in the full Dirac space with $NN$ interactions Bonn A, B, and C, respectively.
Finally, the quadrupole moment of neutron star is calculated under the slow-rotation and small-tidal-deformation approximation.
The equations of state constructed by the RBHF theory in the full Dirac space, together with those by the projection method and momentum-independence approximation, 
conform to universal $I$-Love-$Q$ relations as well.
By combing the tidal deformability from GW170817 and the universal relations from relativistic \textit{ab initio} methods, the moment of inertia of a neutron star with 1.4 solar mass is also deduced as $I_{1.4M_\odot}=1.22^{+0.40}_{-0.25}\times 10^{45}\mathrm{g\ cm^2}$.

\end{abstract}

\maketitle
%\end{CJK*}


%=======================================================================================
\section{Introduction}\label{Sec:Into}
%=======================================================================================

% Importance of neutron stars

Neutron stars are one of the most compact objects in the universe: their central densities can reach as high as 5 to 10 times the saturation density of nuclear matter, $\rho_0 \approx 0.16\ \text{fm}^{-3}$~\cite{Lattimer_2004Science-304-536}, which is far beyond what can be achieved in terrestrial laboratories.
Therefore, neutron stars are ideal laboratories for studying ultradense matter, and have established close connections among nuclear physics, particle physics, and astrophysics.

% Neutron star properties: mass and radius

The astrophysical observations of the global properties of neutron stars provide important constraints for the equation of state (EOS) of dense matter~\cite{Lattimer-2007-Phys.Rep.442.109,Oezel-2016_ARAA54.401,Lattimer-2016_Phys.Rep.621-127,Burgio-2021_PPNP120.103879}, which is the only ingredient needed to unveil
the structure of neutron stars theoretically.
%
%solve the structure equations of neutron stars.
%
%  I did not find the terminology 'the structure equations of neutron star' elsewhere,
%  so I changed the words. 
The high-precision mass measurements of massive neutron stars constitute nowadays one of the most stringent astrophysical constraints on the nuclear EOS; such measurements include 
PSR J1614-2230 ($1.928\pm0.017M_\odot$)~\cite{Demorest-2010_Nature467.1081,Fonseca-2016_ApJ832.167},
PSR J0348+0432 ($2.01\pm0.04M_\odot$)~\cite{Antoniadis-2013_Science340.1233232}, and 
PSR J0740+6620 ($2.08\pm0.07M_\odot$)~\cite{Cromartie-2020_NatureAst4.72,Fonseca_2021-ApJL915.L12}.
Recently, the Neutron star Interior Composition Explorer (NICER) mission reported two independent Bayesian parameter estimations of the mass and equatorial radius of the millisecond pulsar PSR J0030+0451: $1.34_{-0.16}^{+0.15}\ M_\odot$ and $12.71_{-1.19}^{+1.14}\ \text{km}$~\cite{Riley_2019-ApJ887.L21} as well as $1.44_{-0.14}^{+0.15}\ M_\odot$ and $13.02_{-1.06}^{+1.24}\ \text{km}$~\cite{Miller_2019-ApJ887.L24}.
In combination with constraints from radio timing, gravitational wave (GW) observations, and nuclear physics experiments, these posterior distributions have been used to infer the properties of the dense matter EOS (see Ref.~\cite{Riley_2021-ApJ918.L27} and references therein).
Moreover, two independent Bayesian estimations of the radius for the massive millisecond pulsar PSR J0740+6620 have also been reported~\cite{Riley_2021-ApJ918.L27,Miller_2021-ApJL918.L28}.

% Neutron star properties: tidal deformability

Another unique probe for studying the properties of dense matter was extracted from the recent observation of GW signals emitted from a binary neutron star merger, i.e., GW170817~\cite{Abbott2017_PRL119-161101}.
The tidal deformability, which denotes the mass quadrupole moment response of a neutron star to the strong external gravitational field induced by its companion~\cite{Damour_1992-PhysRevD.45.1017,Hinderer_2008-ApJ677.1216H,Flanagan_2008-PhysRevD.77.021502,Damour_2009-PhysRevD.80.084035,Postnikov_2010-PhysRevD.82.024016}, can be inferred from the GW signals.
The limits on the tidal deformability have been widely used to constrain the neutron star radius~\cite{Fattoyev_2018-PhysRevLett.120.172702,Annala_2018-PhysRevLett.120.172703,Most_2018-PhysRevLett.120.261103,Tews_2018-PhysRevC.98.045804}, the asymmetric 
nuclear matter EOS~\cite{Malik_2018-PhysRevC.98.035804,Zhang_2018-ApJ859.90,Tong_2020-PhysRevC.101.035802},  and hence the neutron skin thickness of $^{208}$Pb~\cite{Fattoyev_2018-PhysRevLett.120.172702}. 
%  
% I change the order of theses sentence.
%

% Neutron star properties: moment of inertia

Besides, as rotating objects, the internal structures of neutron stars are strongly constrained by the moment of inertia, which can be determined from the measurements of spin-orbit coupling in double pulsar systems~\cite{Lyne-2004_Science303.1153}. 
Such a measurement of the moment of inertia for neutron star would have crucial implications for delimiting the EOS significantly~\cite{Lattimer_2005-ApJ629.979} and can be used to distinguish neutron stars from quark stars~\cite{Yagi-2014Science341.365}.
Special attention has been attracted by the system PSR J0737-3039~\cite{Burgay-2003_Nature426.531,Lyne-2004_Science303.1153,Kramer_2009-CQG26.073001}, which is the only currently known double pulsar system.
It is hoped that the moment of inertia of the 1.338$M_\odot$ primary component in this system, i.e., PSR J0737-3039A, will be measured eventually to within 10\%~\cite{Kramer_2021-PhysRevX.11.041050}, and could be used to impose new constraints on the EOS~\cite{Lattimer-2021-Universe71.433}.

% Brief to the universal I-Love-Q relations

The global properties of neutron stars, like masses, radii, tidal deformabilities, and moments of inertia are highly sensitive to the EOS for neutron star matter~\cite{YeunhwanLim-2019-PhysRevC.100.035802,JJLi-2019-PhysRevC.100.015809,Tong_2022-AstrophysicsJ930.137}.
Nevertheless, it has been shown~\cite{Yagi-2014Science341.365,Yagi-2013PhysRevD.88.023009} that, for slowly rotating neutron stars, there exist universal relations between the moment of inertia $I$, the tidal deformability $\Lambda$ (or Love number), and the quadrupole moment $Q$ of neutron stars, i.e., the so-called $I$-Love-$Q$ relations, which are approximately independent of the internal composition and the EOS for neutron star matter. 
Wide attention has been attracted by these universal relations (see Ref.~\cite{Yagi-2017Phys.Rep.681.1} for a review).
Although so far the reasons for these universal behaviors are not well understood~\cite{Yagi-2014-PhysRevD.90.063010,Jiang-2020-PhysRevD.101.124006}, attempts have been made to combine the universal relations with GW detections to infer neutron star properties~\cite{Landry_2018-ApJ868.L22,Kumar-2019-PhysRevD.99.123026}.

% Universal I-Love-Q relations from RBHF theory

The robustness of the $I$-Love-$Q$ relations has been extensively studied with EOSs constructed from a variety of nuclear models (see Refs.~\cite{Yagi-2017Phys.Rep.681.1,Wei-JB_2019-JPG46.034001} and references therein), including the relativistic Brueckner-Hartree-Fock (RBHF) theory~\cite{Muether-1987-Phys.Lett.B199.469,Brockmann1990_PRC42-1965,LiGQ-1992-PhysRevC.45.2782,Engvik-1996-ApJ469.794E,Gross-Boelting1999_NPA648-105,Katayama-2015-Phys.Lett.B747.43}.
Since the 1980s the RBHF theory has played an important role in understanding the properties of dense nuclear matter from realistic nucleon-nucleon ($NN$) interactions~\cite{Anastasio1980_PRL45-2096,Anastasio1981_PRC23-2273}.
In the RBHF theory, the single-particle motion of the nucleon in nuclear matter is described with the Dirac equation, where the medium effects are absorbed into the single-particle potentials. 
In principle, the scalar and the vector components of the single-particle potentials should be determined in the full Dirac space~\cite{Nuppenau1989_NPA504-839}, i.e., by considering the positive-energy states (PESs) and negative-energy states (NESs) simultaneously. 
However, to avoid the difficulties induced by NESs, the RBHF calculations are primarily performed in the Dirac space without NESs~\cite{Muether-1987-Phys.Lett.B199.469,Brockmann1990_PRC42-1965,LiGQ-1992-PhysRevC.45.2782,Engvik-1996-ApJ469.794E,Horowitz_1984-Nucl.Phys.B137.287,Gross-Boelting1999_NPA648-105}.

% RBHF theory in the full Dirac space

Recently, a self-consistent RBHF calculation in the full Dirac space was achieved for symmetric nuclear matter (SNM)~\cite{WANG-SB2021_PRC103-054319,WANG-SB2022_PhysRevC.105.054309} and asymmetric nuclear matter (ANM)~\cite{WANG-SB2022_PRC106-L021305}. 
By decomposing the matrix elements of single-particle potential operator in the full Dirac space, the momentum-dependent scalar and vector components of the single-particle potentials are determined uniquely~\cite{WANG-SB2021_PRC103-054319}.
The long-standing controversy about the isospin dependence of the effective Dirac mass in relativistic \textit{ab initio} calculations of ANM is also clarified~\cite{WANG-SB2022_PRC106-L021305}.

% In previous work

The RBHF theory in the full Dirac space has been applied to neutron stars~\cite{WANG-SB2022_PRC106-L021305,Tong_2022-AstrophysicsJ930.137}, where the mass, radius, and tidal deformability are calculated with realistic $NN$ interactions Bonn A, B, and C~\cite{Machleidt1989_ANP19-189}.
The maximum mass of a neutron star is found less than $2.4M_\odot$ and the neutron star radius for $1.4M_\odot$ is predicted about 12 km, which are consistent with the astrophysical observations of massive neutron stars and simultaneous mass-radius estimations by NICER~\cite{Miller_2019-ApJ887.L24}.
The tidal deformabilities for a $1.4M_\odot$ neutron star are predicted as 376, 473, and 459 for the three parametrizations of $NN$ interactions respectively, and all lie in the region $\Lambda_{1.4M_\odot}=190^{+390}_{-120}$ inferred from the revised analysis by LIGO and Virgo Collaborations~\cite{Abbott_2018-PhysRevLett.121.161101}.

% In this work

In this work, we employ the RBHF theory in the full Dirac space to study other global properties of neutron stars, including the gravitational redshift, moment of inertia, and quadrupole moment under the slow-rotation and small-tidal-deformation approximation.
The main focus will be the relation between the moment of inertia and the compactness parameter, as well as the universal $I$-Love-$Q$ relations.
This paper is organized as follows.
In Sec.~\ref{Sec:Theo}, the theoretical framework of the RBHF theory and structure equations for neutron star properties are briefly described.
The obtained results and discussions are presented in Sec.~\ref{Sec:Dis}.
Finally, a summary is given in Sec.~\ref{Sec:Summary}.

%=======================================================================================
\section{Theoretical framework} \label{Sec:Theo}
%=======================================================================================

\subsection{The relativistic Brueckner-Hartree-Fock theory}

In the RBHF calculations, one of the most important procedure is the self-consistent determination of the single-particle potential operator $\mathcal{U}$ of the nucleons, which is generally divided into scalar and vector components~\cite{Serot_1986-ANP}
\begin{equation}\label{eq:SPP}
  \mathcal{U}(\bm{p}) = U_{S}(p)+ \gamma^0U_{0}(p) + \bm{\gamma\cdot\hat{p}}\,U_{V}(p).
\end{equation}
Here $\hat{\bm{p}}=\bm{p}/p$ is the unit vector parallel to the momentum $\bm{p}$. 
The quantities $U_S(p)$, $U_0(p)$, and $U_V(p)$ are the scalar potential, the timelike part and the spacelike part of the vector potential.

In principle, the scalar and the vector components of the single-particle potentials can only be determined uniquely in the full Dirac space.
However, to avoid the numerical difficulties in the full Dirac space, different approximations are proposed to extract the single-particle potentials in the Dirac space without NESs.
The momentum-independence approximation~\cite{Brockmann1990_PRC42-1965} assumes that the single-particle potentials are independent of the momentum, and the spacelike component of the vector potential, $U_{V}$, is negligible.
The scalar potential $U_{S}$ and the timelike part of the vector potential, $U_{0}$, are then extracted from the single-particle potential energies at two selected momenta.
In the projection method~\cite{Gross-Boelting1999_NPA648-105}, the effective $NN$ interaction $G$ matrix, which is obtained by solving the in-medium scattering equation, is projected onto a complete set of five Lorentz invariant amplitudes, from which the single-particle potentials are calculated analytically.
However, the choice of these Lorentz invariant amplitudes is not unique.

Only by decomposing the matrix elements of $\mathcal{U}$ in the full Dirac space, can the Lorentz structure and momentum dependence of single-particle potentials be uniquely determined~\cite{Nuppenau1989_NPA504-839}.
The theoretical framework for the RBHF theory in the full Dirac space has been described in detail in Ref.~\cite{WANG-SB2021_PRC103-054319} for SNM and Ref.~\cite{Tong_2022-AstrophysicsJ930.137} for ANM.
In this work this method is used to construct the EOS of neutron star matter, which is regarded as beta equilibrium nuclear matter consisting of protons, neutrons, electrons, and muons~\cite{Krastev_2006PRC-74-025808}.
Using the relativistic Bonn A potential~\cite{Machleidt1989_ANP19-189}, the RBHF theory in the full Dirac space for nuclear matter is applicable for density in the range 0.08-0.57 $\text{fm}^{-3}$.
For lower density in the crust of a neutron star, the EOS introduced with the Baym-Bethe-Pethick (BBP)~\cite{BBP-1971_Nucl.Phys.A175.225} and Baym-Pethick-Sutherland (BPS) model~\cite{BaymPethickSutherland-1971-ApJ.170.299B} is used.
For higher density, we follow the strategy proposed in Ref.~\cite{Rhoades1974_PRL32-324} and applied in Refs.~\cite{Gandolfi2012_PRC85-032801,WANG-SB2022_PRC106-L021305}, where the neutron-star matter EOS above a critical density $\rho_c=0.57\ \text{fm}^{-3}$ is replaced with the maximally stiff or causal one, which predicts the most rapid increase of pressure with energy density without violating the causality limit.

\subsection{Mass, radius, gravitational redshift, and tidal deformability}

The stable configurations of a cold, spherically symmetric, and nonrotating neutron star can be obtained from the Tolman-Oppenheimer-Volkov (TOV) equations~\cite{Oppenheimer1939_PR55-374,Tolman1939_PR55-364}.
Adopting natural units $G=c=1$, the TOV equations are given by
\begin{subequations}\label{eq:TOV}
  \begin{align}
    \frac{dP(r)}{dr}=&\ -\frac{[P(r)+\mathcal{E}(r)][M(r)+4\pi r^3P(r)]}{r[r-2M(r)]}, \\
	\frac{dM(r)}{dr} =&\ 4\pi r^2\mathcal{E}(r),
  \end{align}
\end{subequations}
where $P(r)$ is the pressure at neutron star radius $r$, $M(r)$ is the total neutron star mass inside a sphere of radius $r$, and $\mathcal{E}(r)$ is the total energy density.
These differential equations can be solved numerically with a given central pressure $P_c$ and $M(0)=0$. 
The quantity $R$ for $P(R)=0$ denotes the radius of the neutron star, and $M(R)$ is its mass.
The gravitational redshift which relates the mass of the neutron star to its radius is defined as
\begin{equation} 
	z = \left( 1 - 2M/R\right)^{-1/2} - 1.
\end{equation}
Since the radius of the neutron star is harder to observe relative to its mass, the simultaneous measurements of the mass and the gravitational redshift would provide a clear radius determination.

The tidal deformability is defined as
\begin{equation} 
	\Lambda = \frac{2}{3}k_2 C^{-5}.
\end{equation}
where $C=M/R$ is the compactness parameter. The second Love number $k_2$~\cite{Hinderer_2008-ApJ677.1216H,Hinderer_2010-PhysRevD.81.123016} is calculated by
\begin{equation} 
  \begin{split}
    k_2 = &\ \frac{8C^5}{5}(1-2C)^2 [2-y_R + 2C(y_R-1)] \times \left\{ 6C[2-y_R+C(5y_R-8)] \right.\\
      &\ + 4C^3[13-11y_R+C(3y_R-2)+2C^2(1+y_R)] \\
      &\ \left. + 3(1-2C)^2[2-y_R+2C(y_R-1)]\ln (1-2C)  \right\}^{-1},
  \end{split} 
\end{equation}
where $y_R=y(R)$ is the solution of the following nonlinear, first-order differential equation
\begin{equation}\label{eq:y}
  r\frac{dy(r)}{dr} + y^2(r) + F(r)y(r) + r^2Q(r) =0.
\end{equation}
Here the two functions $F(r)$ and $Q(r)$ depend on the known mass, radius, pressure, and energy density profiles of the star:
\begin{subequations}\label{eq:TOV}
  \begin{align}
    F(r) =&\ \left[ 1-\frac{2M(r)}{r}\right]^{-1} \left\{ 1-4\pi r^2[\mathcal{E}(r)-P(r)]\right\},\\
    Q(r) =&\ \left\{ 4\pi \left[ 5\mathcal{E}(r)+9P(r)+\frac{\mathcal{E}(r)+P(r)}{\partial P/\partial \mathcal{E}}\right]-\frac{6}{r^2} \right\}
              \times \left[ 1-\frac{2M(r)}{r}\right]^{-1} \nonumber \\
          &\ - \left[ \frac{2M(r)}{r^2}+8\pi r P(r)\right]^2  \times \left[ 1-\frac{2M(r)}{r}\right]^{-2} 
  \end{align}
\end{subequations}
The differential equation~\eqref{eq:y} for $k_2$ can be solved together with the TOV equations and the initial condition $y(0)=2$.

\subsection{The moment of inertia}

The moment of inertia is calculated under the slow-rotation approximation pioneered by Hartle and Thorne~\cite{Hartle-1967ApJ150.1005,Hartle-1968-ApJ153.807H}, where the frequency $\Omega$ of a uniformly rotating neutron star is far smaller than the Kepler frequency at the equator:
\begin{equation} 
	\Omega \ll \Omega_{\text{max}} \simeq \sqrt{M/R^3}.
\end{equation}
In the slow-rotation approximation the moment of inertia of a uniformly rotating, axially symmetric neutron star is given by the following expression~\cite{Fattoyev-2010-PhysRevC.82.025810}:
\begin{equation}
	I = \frac{8\pi}{3} \int^R_0 r^4 e^{-\nu(r)} \frac{\bar{\omega}(r)}{\Omega} \frac{\mathcal{E}(r)+P(r)}{\sqrt{1-2M(r)/r}} dr.
\end{equation}
The quantity $\nu(r)$ is a radially dependent metric function and is defined as
\begin{equation}
	\nu(r) = \frac{1}{2}\ln \left( 1- \frac{2M}{R}\right) - \int^R_r \frac{M(x)+4\pi x^3 P(x)}{x^2[1-2M(x)/x]}dx.
\end{equation}
The frame-dragging angular velocity $\bar{\omega}$ is usually obtained by the dimensionless relative frequency $\tilde{w}\equiv \bar{\omega}/\Omega$, which satisfies the following second-order differential equation:
\begin{equation}
  \frac{d}{dr}\left[ r^4 j(r)\frac{d\tilde{\omega}(r)}{dr}\right]  + 4r^3 \frac{dj(r)}{dr} \tilde{\omega}(r) = 0,
\end{equation}
where $j(r) = e^{-\nu(r)} \sqrt{1-2M(r)/r}$ for $r\leq R$. 
The relative frequency $\tilde{\omega}(r)$ is subject to the following two boundary conditions
\begin{subequations}
  \begin{align}
    \tilde{\omega}'(0) =&\ 0,\\
    \tilde{\omega}(R) + \frac{R}{3}\tilde{\omega}'(R) =&\ 1.
  \end{align} 
\end{subequations}
It should be noted that under the slow-rotation approximation the moment of inertia does not depend on the stellar frequency $\Omega$.

\subsection{The quadrupole moment}

It has been shown~\cite{Yagi-2014Science341.365,Yagi-2013PhysRevD.88.023009} that there exist universal relations between the moment of inertia, the Love number, and the quadrupole moment of neutron stars.
Physically, the moment of inertia quantifies how fast a neutron star can spin for a fixed angular momentum, the quadrupole moment describes how much a neutron star is deformed away from sphericity due to rotation, and the Love number characterizes how easily a neutron star can be deformed due to an external tidal field.
These quantities can be computed by numerically solving for the interior and exterior gravitational field of a neutron star in a slow-rotation~\cite{Hartle-1967ApJ150.1005,Hartle-1968-ApJ153.807H} approximation and in a small-tidal-deformation approximation~\cite{Hinderer_2008-ApJ677.1216H,Hinderer_2010-PhysRevD.81.123016}.
In this work the quadrupole moment is calculated by following the detailed instructions described in Ref.~\cite{Yagi-2013PhysRevD.88.023009}.
In order to investigate the universal $I$-Love-$Q$ relations, the following dimensionless quantities are introduced:
\begin{equation}
	\bar{I} \equiv \frac{I}{M^3},\qquad \bar{Q} \equiv - \frac{QM}{(I\Omega)^2}.
\end{equation}

% %=======================================================================================
% \section{Numerical details}\label{Sec:Num}
% %=======================================================================================

%=======================================================================================
\section{Results and discussions}\label{Sec:Dis}
%=======================================================================================

% Fig. 1: speed of sound and mass-radius relation

\begin{figure}[htbp]
  \centering
  \includegraphics[width=14.0cm]{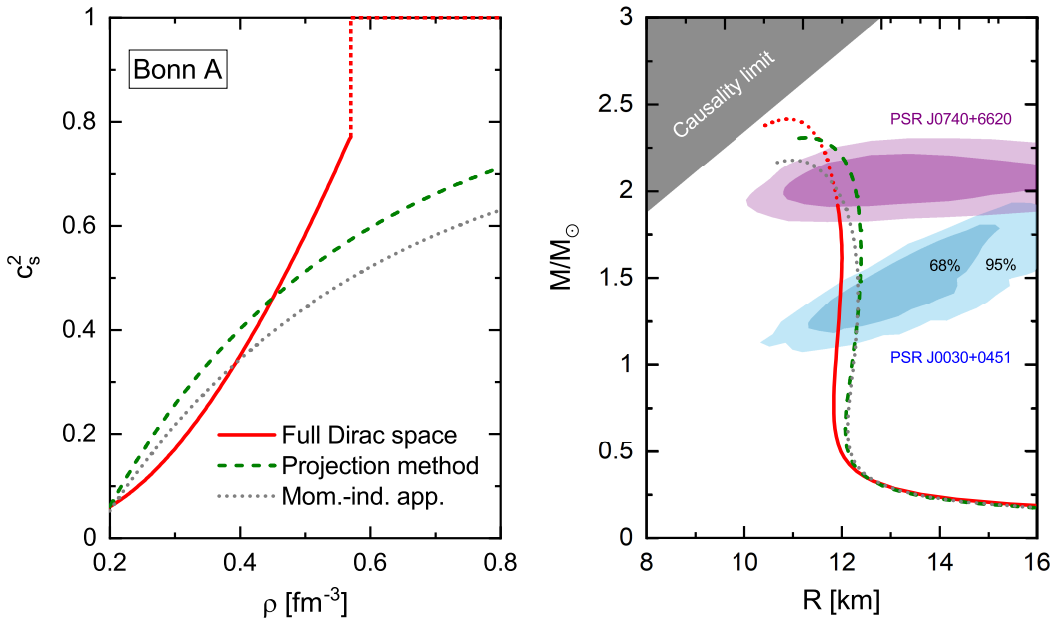}
  \caption{(Color online) The speed of sound squared for neutron star matter as a function of density (left) and the neutron star mass-radius relation (right) obtained by the RBHF theory with the potential Bonn A in the full Dirac space (red solid line), in comparison with those obtained by the projection method (green dashed line) and the momentum-independence approximation (gray dotted line). The dark (light) blue and purple regions indicate the 68\% (95\%) confidence intervals constrained by the NICER analysis of PSR J0030+0451~\cite{Miller_2019-ApJ887.L24} and PSR J0740+6620~\cite{Miller_2021-ApJL918.L28}. The red dotted line corresponds to the maximally stiff EOS.}
  \label{Fig1}
\end{figure}

In the left panel of Fig.~\ref{Fig1}, the speed of sound squared for neutron star matter from the RBHF theory with potential Bonn A is depicted as a function of density.
The results calculated with the projection method and momentum-independence approximation show slowly increasing tendencies with the increase of density. 
In the full Dirac space, the speed of sound squared increases quickly with the density, and reaches 0.77 at $0.57\ \text{fm}^{-3}$.
Above this density, the amplitude of scalar potential $U_S$ exceeds the nucleon rest mass, and the RBHF iteration in the full Dirac space is very difficult to achieve.
This fact might be related to the extension of the Bonn potential to the full Dirac space, which was determined with PESs only. 
When the central density of a neutron star is fixed at $\rho_c=0.57\ \text{fm}^{-3}$, our calculations could support a neutron star with mass equal to $1.9M_\odot$. 
To further explore the maximum mass of a neutron star from the RBHF theory in the full Dirac space, we continue with an EOS where the speed of sound is equal to the speed of light~\cite{Rhoades1974_PRL32-324,Gandolfi2012_PRC85-032801,WANG-SB2022_PRC106-L021305} (red dotted line). 
This would provide an upper bound on the maximum mass of a neutron star.

Based on the EOS from the RBHF theory, the mass-radius relations of a neutron star can be calculated from the TOV equations, which are shown in the right panel of Fig.~\ref{Fig1}.
The 68\% and 95\% contours of the joint probability density distribution of the mass and radius of PSR J0030+0451~\cite{Miller_2019-ApJ887.L24} and PSR J0740+6620~\cite{Miller_2021-ApJL918.L28} from the NICER analysis are also shown.
It can be found that the results obtained in the full Dirac space, with the projection method, and with the momentum-independence approximation are consistent with the recent constraints by NICER. 
The maximum masses $M_\text{max}$ predicted by the three methods are $2.43M_\odot$, $2.31M_\odot$, and $2.18M_\odot$ respectively, which are consistent with the available astrophysical constraints from massive neutron star observations, such as PSR J1614-2230 ($1.928\pm0.017M_\odot$)~\cite{Demorest-2010_Nature467.1081,Fonseca-2016_ApJ832.167}, PSR J0348+0432 ($2.01\pm0.04M_\odot$)~\cite{Antoniadis-2013_Science340.1233232}, and PSR J0740+6620 ($2.08\pm0.07M_\odot$)~\cite{Cromartie-2020_NatureAst4.72,Fonseca_2021-ApJL915.L12}.
The radii $R_{1.4M_\odot}$ of a $1.4M_\odot$ neutron star from the three methods are $11.97\ \text{km}$, $12.38\ \text{km}$, and $12.35\ \text{km}$, respectively.

% Tab. 1: nuclear matter properties and neutron star properties from the RBHF theory

\begin{table}[htbp]
  \centering
  \caption{Neutron star properties and nuclear matter properties at saturation density calculated by the RBHF theory in the full Dirac space with potential Bonn A, in comparison with the results obtained by the RBHF calculation with the projection method and the momentum-independence approximation.}
  \begin{tabular}{cccccccc}
    \hline\hline
    
    \multirow{2}{*}{Model} & $M_{\text{max}}$ & $R_{1.4M_\odot}$ & $\rho_{1.4M_\odot}$ & $\rho_0$    & $E/A$ & $E_{\mathrm{sym}}$ & $L$\\
    					   & ($M_\odot$)      & (km)             & (fm$^{-3}$)         & (fm$^{-3}$) & (MeV) & (MeV)              & (MeV) \\
    \hline
    Full Dirac space  & 2.43 & 11.97 & 0.46 & 0.188 & -15.40 & 33.1 & 65.2 \\
    Projection method & 2.31 & 12.38 & 0.42 & 0.179 & -16.15 & 34.7 & 68.8 \\
	Mom.-ind. app.    & 2.18 & 12.35 & 0.43 & 0.178 & -15.36 & 33.2 & 67.3 \\
    \hline

  \end{tabular}
  \label{tab1}
\end{table}

In Table \ref{tab1} we summarize the maximum mass of a neutron star and the radius and the central density for a $1.4M_\odot$ neutron star obtained by the RBHF theory in the full Dirac space, together with those obtained with the projection method and momentum-independence approximation. 
The smallest value for $R_{1.4M_\odot}$ found in the full Dirac space corresponds to the softest EOS for neutron star matter below a density of about 0.4 fm$^{-3}$.
This fact can be further related to nuclear matter properties, which are also summarized in Tab.~\ref{tab1}, including the binding energy per nucleon $E/A$, the symmetry energy $E_\text{sym}$, and its slope $L$.
It is found that the symmetry energy and its slope at the saturation density obtained in the full Dirac space are the smallest among the three methods. This explains the reason for smallest $R_{1.4M_\odot}$ and shows how important it is to take both the PESs and NESs into account.

% Fig. 2: Gravitational redshift with mass and radius

\begin{figure}[htbp]
  \centering
  \includegraphics[width=14.0cm]{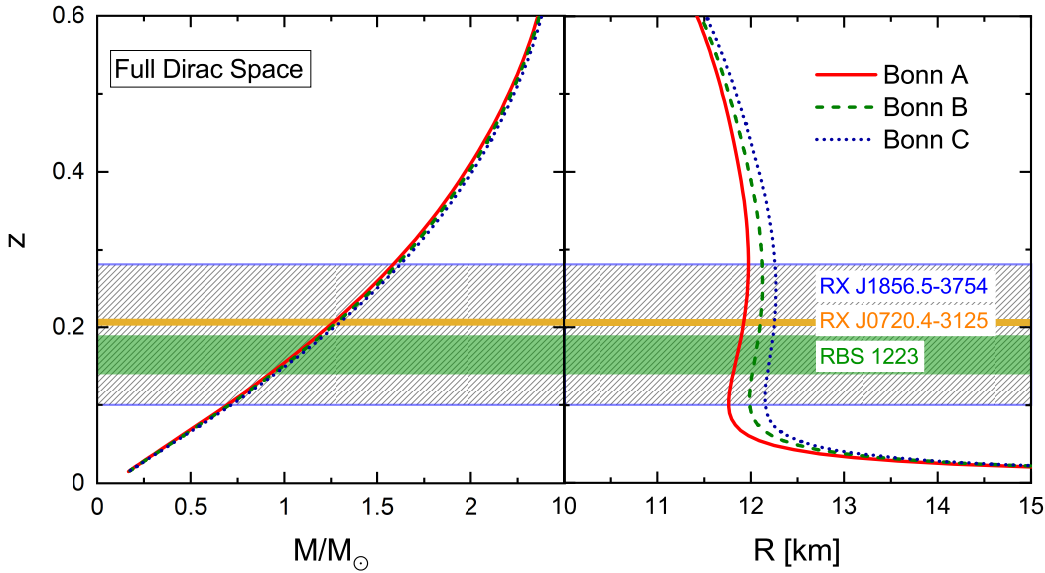}
  \caption{(Color online) The relation between the gravitational redshift $z$ of a neutron star with its mass $M$ (left) and radius $R$ (right) obtained by the RBHF theory in the full Dirac space with the potentials Bonn A, B, and C. Astrophysical observation for the gravitational redshift of isolated neutron stars RBS 1223~\cite{Hambaryan_2014-JPCS}, RX J1856.5-3754~\cite{Hambaryan_2014-JPCS}, and RX J0720.4-3125~\cite{Hambaryan_2017-AA601.A108} are also shown.}
  \label{Fig2}
\end{figure}

Figure~\ref{Fig2} shows the relations between the gravitational redshift $z$ of a neutron star with its mass $M$ and radius $R$ obtained by the RBHF theory in the full Dirac space with the potentials Bonn A, B, and C.
It can be seen that the $z$-$M$ relation is not strongly affected by the $NN$ interactions.
With the increase of mass, the gravitational redshift shows a monotonically increasing behavior.
A backbending phenomenon is found for the gravitational redshift with the decrease of radius, similar to the case for $M$-$R$ relations as shown in Fig.~\ref{Fig1}.

% Tab. 2: Gravitational redshift of isolated neutron stars

\begin{table}[htbp]
  \centering
  \caption{The masses of isolated neutron star RBS 1223~\cite{Hambaryan_2014-JPCS}, RX J1856.5-3754~\cite{Hambaryan_2014-JPCS}, and RX J0720.4-3125~\cite{Hambaryan_2017-AA601.A108} predicted by combining gravitational redshift measurements and the RBHF theory in the full Dirac space with potential Bonn A. 
  The radii are also given. The uncertainties in the last two columns are all from the gravitational redshift measurements.}
  \setlength{\tabcolsep}{3mm}{
  \begin{tabular}{cccc}
    \hline\hline
    System & Gravitational redshift $z$  & Mass\ $M[M_\odot]$  & Radius $R[\text{km}]$ \\
    \hline
    RBS 1223~\cite{Hambaryan_2014-JPCS}              & $0.16^{+0.03}_{-0.02}$      & $1.03^{+0.15}_{-0.11}$    & $11.85^{+0.05}_{-0.04}$\\
    RX J1856.5-3754~\cite{Hambaryan_2014-JPCS}       & $0.22^{+0.06}_{-0.12}$      & $1.33^{+0.25}_{-0.64}$    & $11.94^{+0.03}_{-0.18}$ \\
    RX J0720.4-3125~\cite{Hambaryan_2017-AA601.A108} & $0.205^{+0.006}_{-0.003}$   & $1.258^{+0.028}_{-0.014}$ & $11.922^{+0.008}_{-0.004}$\\  
    \hline\hline
  \end{tabular}}	
  \label{Tab2}
\end{table}

The clear one-to-one correspondence relation for gravitational redshift and mass established in the left panel in Fig.~\ref{Fig2} can be used to infer the mass of a isolated neutron star,  when the observation of the gravitational redshift is provided.
In Fig.~\ref{Fig2}, astrophysical observations of gravitational redshift for isolated neutron stars RX J0720.4-3125~\cite{Hambaryan_2017-AA601.A108}, 
RBS 1223~\cite{Hambaryan_2014-JPCS}, and RX J1856.5-3754~\cite{Hambaryan_2014-JPCS} are shown as shadow bands.
The predicted masses by combining these observations and the theoretical calculations from the RBHF theory in the full Dirac space with Bonn A potential are listed in the third column of Table \ref{Tab2}.
The uncertainties are from the gravitational redshift measurements.
In Table \ref{Tab2} the corresponding radii are also shown in the last column.

% Fig. 3: Universial relation of momenta of inertia

\begin{figure}[htbp]
  \centering
  \includegraphics[width=10.0cm]{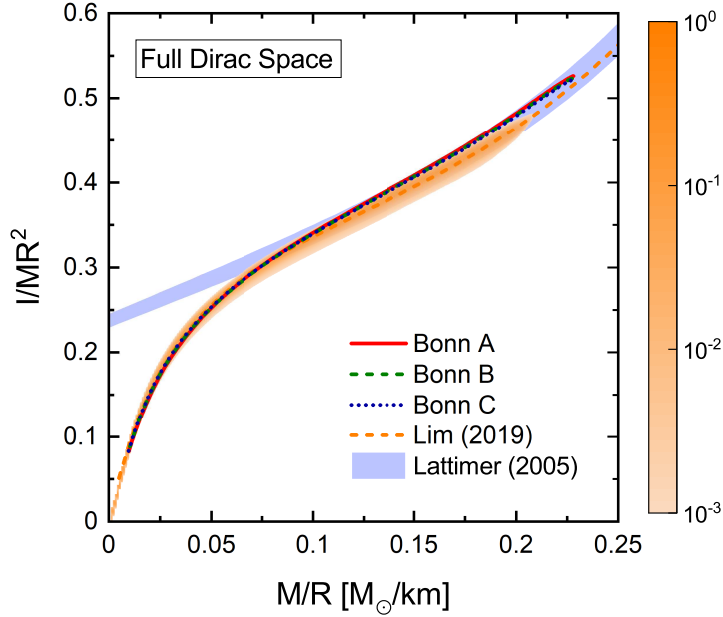}
  \caption{(Color online) The ratio of the moment of inertia $I$ to $MR^2$ as a function of the compactness parameter $M/R$ obtained with the RBHF theory in the full Dirac space with $NN$ interactions Bonn A, B, and C.
  The probability distribution (orange) from the Bayesian analysis together with the fitted (orange) line in Ref.~\cite{YeunhwanLim-2019-PhysRevC.100.035802} are shown for comparison.
  The empirical (purple) band from Ref.~\cite{Lattimer_2005-ApJ629.979} is also shown.
  See the text for details.}
  \label{Fig3}
\end{figure}

In Fig.~\ref{Fig3} we display the ratio of the moment of inertia $I$ to $MR^2$ as a function of the compactness parameter $M/R$ obtained by the RBHF theory in the full Dirac space with $NN$ interactions Bonn A, B, and C.
Lattimer et al.~\cite{Lattimer_2005-ApJ629.979} showed that, in the absence of phase transition and other effects that strongly soften the EOS at supranuclear densities, there is a relatively unique relation between the quantity $I/MR^2$ and $M/R$:
\begin{equation}
	I/MR^2 \simeq (0.237\pm 0.008) (1+2.844C + 18.91C^4).
\end{equation}
This relation is shown as the purple band in Fig.~\ref{Fig3}.
It is found that our results are consistent with the universal relations obtained in Ref.~\cite{Lattimer_2005-ApJ629.979} for the range where $M/R>0.08\ M_\odot/\text{km}$.
The derivation for smaller compactness is unimportant, since the observational evidence for neutron star masses and radii lie in the ranges of $1.2\ M_\odot\leq M\leq 2.2\ M_\odot$ and $9\ \text{km}\leq R\leq 15\ \text{km}$ respectively, which leads to the range of compactness as $0.08\ M_\odot/\text{km} < M/R < 0.24\ M_\odot/\text{km}$.
Lim and collaborators~\cite{YeunhwanLim-2019-PhysRevC.100.035802} have investigated neutron star moments of inertia from Bayesian posterior probability distributions of the nuclear EOSs that incorporate information from microscopic many-body theory and empirical data of finite nuclei.
The probability distribution for $I/MR^2$ is shown as  in Fig.~\ref{Fig3}.
They found that over the entire range of neutron star compactness $0< C\leq 0.25\ M_\odot/\text{km}$, their results can be well fitted with the following formula:
\begin{equation}\label{eq:IQbar}
	I/MR^2 = \frac{C+27.178C^4}{0.0871+2.183C}.
\end{equation}
This result is also shown as the orange dashed line in Fig.~\ref{Fig3}.
It can be seen that our results are very close to that obtained by Lim et al., especially for the neutron stars with small compactness.

% Tab. 3: momenta of inertia of J0737-3039A

\begin{table}[htbp]
  \centering
  \caption{The momenta of inertia for $1.338\ M_\odot$ pular PSR J0737-3039A predicted by the RBHF theory in the full Dirac space with potentials Bonn A, B, and C.
  The results obtained with the projection method~\cite{Gross-Boelting1999_NPA648-105,VanDalen2004_NPA744-227} and momentum-independence approximation (Mom.-ind. approx.)~\cite{Brockmann1990_PRC42-1965} are also shown. For comparison, the values obtained with nonrelativistic \textit{ab initio} calculation~\cite{Morrison_2004-ApJ617.L135}, predicted with the tidal deformability in GW170817 combing universal relations~\cite{Landry_2018-ApJ868.L22} as well as that inferred from Bayesian analysis (95\% credibility)~\cite{YeunhwanLim-2019-PhysRevC.100.035802} are also shown. The last column is the corresponding radius.}
  \setlength{\tabcolsep}{1mm}{
  \begin{tabular}{cccc}
    \hline\hline

    Model  &  Potential & $I_{1.338M_\odot}~[10^{45}\ \mathrm{g~cm^2}]$ & $ R_{1.338M_\odot}$~[km]\\
    \hline
                     & A & 1.356 & 11.94 \\
    Full Dirac space & B & 1.381 & 12.11 \\
                     & C & 1.407 & 12.26 \\
    \hline     
    				       & A & 1.440 & 12.34 \\
    Projection method      & B & 1.465 & 12.49 \\
    				       & C & 1.487 & 12.60 \\
    \hline
    				  & A & 1.431 & 12.32 \\
    Mom.-ind. approx. & B & 1.452 & 12.46 \\
    				  & C & 1.471 & 12.57 \\
    \hline
    Variational calculation (APR)~\cite{Morrison_2004-ApJ617.L135}  & AV18 $+\ \delta v$ $+$ UIX* & 1.24                   & 11.56\\
    GW170817 $+$ universal relations~\cite{Landry_2018-ApJ868.L22}  &                             & $1.15^{+0.38}_{-0.24}$ & \\
    Bayesian analysis~\cite{YeunhwanLim-2019-PhysRevC.100.035802}   &                             & $1.36^{+0.15}_{-0.32}$ & $12.2^{+0.7}_{-1.9}$\\
    \hline\hline
  \end{tabular}}
  \label{Tab3}
\end{table}

%. interest of J0737-3039

Although the ratio of the neutron star moment of inertia $I$ to $MR^2$ has a universal function of the compactness parameter $M/R$, the moment of inertia itself depends sensitively on the neutron star's internal structure.
It has been suggested~\cite{Lattimer_2005-ApJ629.979} that a measurement accuracy of 10\% for $I$ is sufficient to place strong constraints on the EOS.

In Table \ref{Tab3}, we show the momenta of inertia $I_{1.338M_\odot}$ and radius $R_{1.338M_\odot}$ for PSR J0737-3039A predicted by the RBHF theory in the full Dirac space with three parametrizations for $NN$ interactions.
The results obtained by the projection method~\cite{Gross-Boelting1999_NPA648-105,VanDalen2004_NPA744-227} and momentum-independence approximation~\cite{Brockmann1990_PRC42-1965} are also shown.
The RBHF theory in the full Dirac space leads to minimum values compared to the approximations in the Dirac space without NESs.
This is understandable since the RBHF theory in the full Dirac space gives the minimum radius of a neutron star for the fixed canonical mass, as shown in Table \ref{tab1}.

The moment of inertia for PSR J0737-3039A predicted by the RBHF theory in the full Dirac space with Bonn A is $1.356\times 10^{45}\ \mathrm{g~cm^2}$, which is very close to the most probable value $1.36\times 10^{45}\ \mathrm{g~cm^2}$ obtained from Bayesian analysis (95\% credibility)~\cite{YeunhwanLim-2019-PhysRevC.100.035802}.
The result from the nonrelativistic \textit{ab initio} variational calculations~\cite{Akmal_1998PRC-58-1804} is also shown in Table \ref{Tab3}, where the Argonne $v18$ interaction (AV18)~\cite{Wiringa1995_PRC51-38} is used, together with the relativistic boost corrections to the two-nucleon interaction as well as three-nucleon interactions modeled with the Urbana force~\cite{Pudliner1995_PRL74-4396}.
The nonrelativistic \textit{ab initio} calculation leads to a moment of inertia smaller that what we obtain, similarly to the case for radius.
In Ref.~\cite{Landry_2018-ApJ868.L22}, by using well-known universal relations among neutron star observables, the reported 90\% credible bound on the tidal deformability $\Lambda_{1.4M_\odot}=190^{+390}_{-120}$ from GW170817~\cite{Abbott_2018-PhysRevLett.121.161101} has been translated into a direct constraint on the moment of inertia of PSR J0737-3039A, giving $I_{1.338M_\odot}=1.15^{+0.38}_{-0.24}\times 10^{45}\ \mathrm{g~cm^2}$. 
It can be seen that the results with the three methods for RBHF theory are consistent with this constraint.

% Fig. 4: Universial relation of I-Love and Q-Love

\begin{figure}[htbp]
  \centering
  \includegraphics[width=7.5cm]{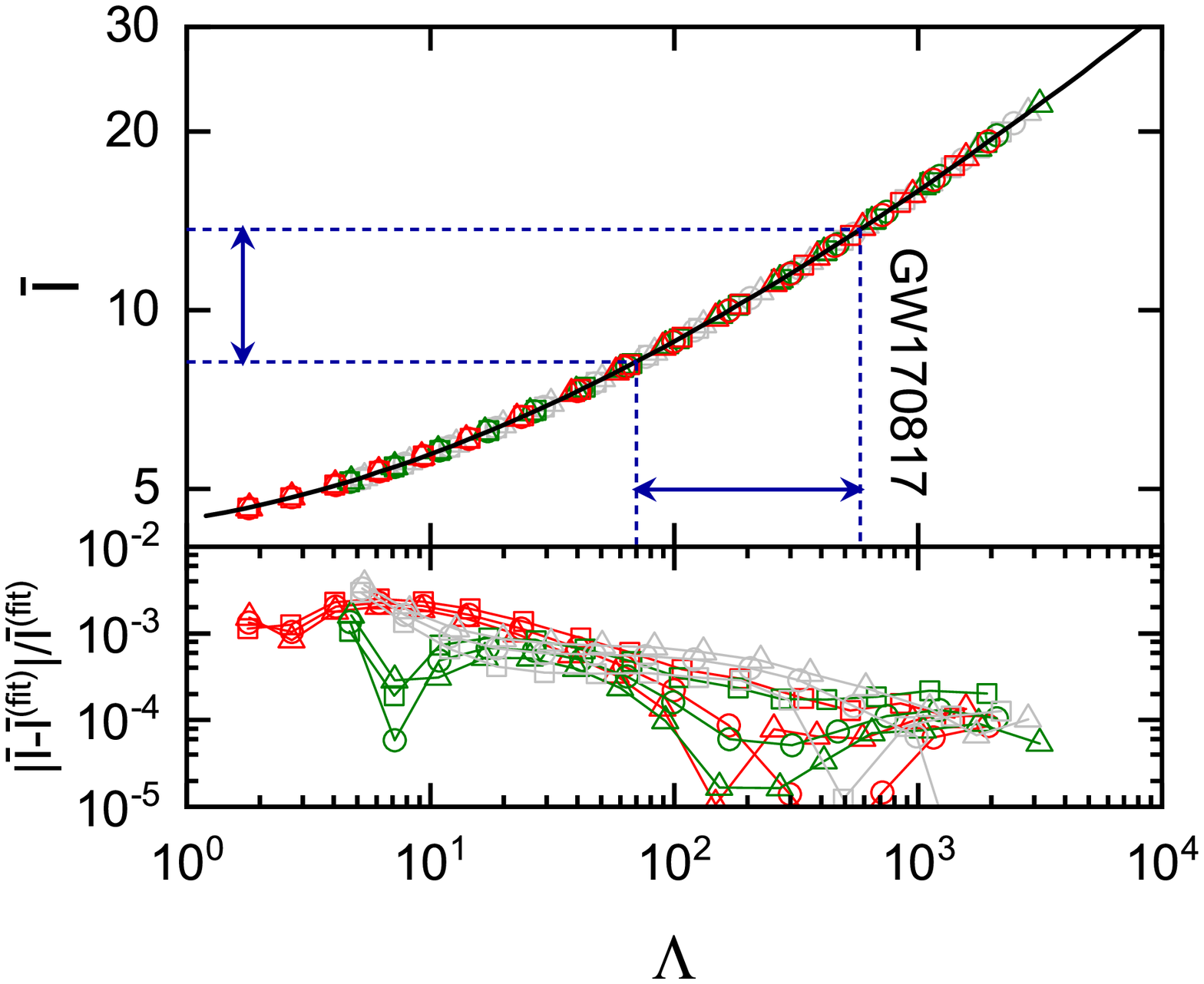}\qquad 
  \includegraphics[width=7.5cm]{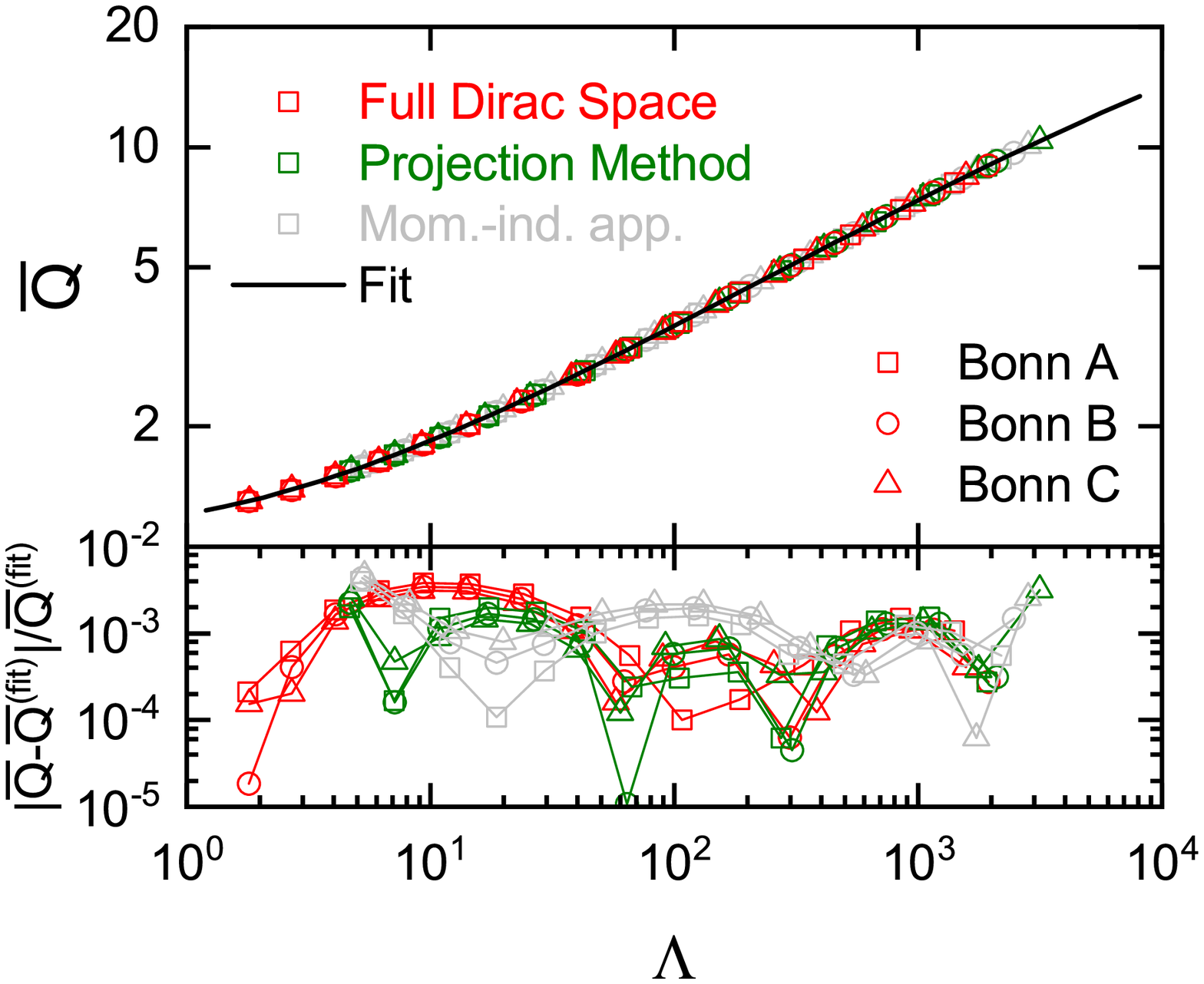}
  \caption{(Color online) Top: The $I$-Love (left) and $Q$-Love (right) relations for slowly-rotating neutron stars with EOSs obtained by the RBHF theory in the full Dirac space, projection method, and momentum-independence approximations with potentials Bonn A, B, and C.
  The solid curves show the fit by using Eq.~\eqref{eq:Formula-I-Love-Q}.
  Bottom: Absolute fractional difference from the fit.}
  \label{Fig4}
\end{figure}

\begin{figure}[htbp]
  \centering
  \includegraphics[width=7.5cm]{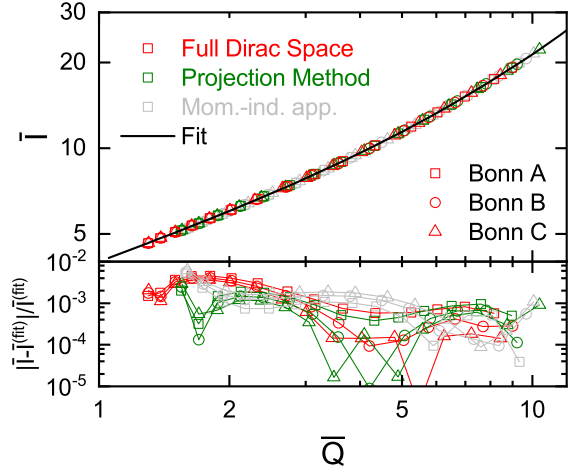}
  \caption{(Color online) The same as the left panel in Fig.~\ref{Fig4}, but for $I$-$Q$ relations.}
  \label{Fig5}
\end{figure}

Let us now compare the EOSs obtained by the relativistic \textit{ab initio} calculations, i.e., the RBHF theory in the full Dirac space, projection method, and momentum-independence approximation with Bonn potentials, to the universal $I$-Love-$Q$ relations. 
The $I$-Love as well as $Q$-Love relations and $I$-$Q$ relations are shown in the top panels of Figs.~\ref{Fig4} and \ref{Fig5}, respectively.
The dimensionless moment of inertia $\bar{I}$ and dimensionless quadrupole moment $\bar{Q}$ are defined in Eq.~\eqref{eq:IQbar}.
A single parameter along the curve is the mass or compactness, which increases to the left of the plots.
Similarly to Ref.~\cite{Yagi-2017Phys.Rep.681.1}, we only show data with the mass of an isolated, nonrotating configuration in the range $1M_\odot<M<M_{\text{max}}$ with $M_{\text{max}}$ representing the maximum mass for such a configuration.
One observes that the universal relations hold very well. 
Since the relations are insensitive to EOS, one can construct a single fit (black solid curves) given by~\cite{Yagi-2013PhysRevD.88.023009,Yagi-2017Phys.Rep.681.1}
\begin{equation}\label{eq:Formula-I-Love-Q}
	\ln y_i = a_i + b_i \ln x_i + c_i (\ln x_i)^2 + d_i (\ln x_i)^3 + e_i (\ln x_i)^4,
\end{equation}
where coefficients are listed in Table \ref{Tab4}.
These coefficients are very close to that in Ref.~\cite{Yagi-2017Phys.Rep.681.1}, where a large number of EOSs are considered.
The bottom panels of Figs.~\ref{Fig4} and \ref{Fig5} show the absolute fractional difference between all the data and the fit, which is less than $1\%$ in the whole range.

The universal relations between $\bar{I}$ and $\Lambda$ allows one to extract the momenta of inertia of a neutron star with $1.4$ solar mass, $\bar{I}_{1.4M_\odot}$, from the tidal deformability $\Lambda_{1.4M_\odot}$ from GW170817.
The revised analysis from LIGO and Virgo Collaborations, $\Lambda_{1.4M_\odot}=190^{+390}_{-120}$~\cite{Abbott_2018-PhysRevLett.121.161101}, corresponds to $\bar{I}_{1.4M_\odot}=10.30^{+3.39}_{-2.10}$ as shown in the left panel of Fig.~\ref{Fig4}. From $\bar{I}_{1.4M_\odot}$ and the relation $\bar{I}=I/M^3$ we obtain $I_{1.4M_\odot}=1.22^{+0.40}_{-0.25}\times 10^{45}\mathrm{g\ cm^2}$.
These values are consistent with the results $\bar{I}_{1.4M_\odot}=11.10^{+3.64}_{-2.28}$ and $I_{1.4M_\odot}=1.15^{+0.38}_{-0.24}\times 10^{45}\mathrm{g\ cm^2}$ in Ref.~\cite{Landry_2018-ApJ868.L22}, where the $I$-Love relation is obtained by using a large set of candidate neutron star EOSs based on relativistic mean-field and Skyrme-Hartree-Fock theory.

% Fig. 5: Universial relation of I-Q

% Tba. 3: numerical coefficients for I-Love-Q

\begin{table}[htbp]
  \centering
  \caption{Numerical coefficients for the fitting formula of the $I$-Love, $I$-$Q$, and $Q$-Love relations given in Eq.~\eqref{eq:Formula-I-Love-Q}. }
  \setlength{\tabcolsep}{5mm}{
  \begin{tabular}{ccccccc}
    \hline\hline
    $y_i$ & $x_i$ & $a_i$ & $b_i$ & $c_i$ & $d_i$ & $e_i$ \\
    \hline
    $\bar{I}$ & $\Lambda$ & 1.493  & 0.06409 & 0.02104 & $-5.381\times 10^{-4}$ & $1.957\times 10^{-6}$ \\
    $\bar{I}$ & $\bar{Q}$ & 1.387  & 0.5722  & 0.01043 & $0.02342$              & $6.245\times 10^{-4}$ \\
    $\bar{Q}$ & $\Lambda$ & 0.1899 & 0.09937 & 0.04380 & $-3.430\times 10^{-3}$ & $7.054\times 10^{-5}$ \\
    \hline\hline
  \end{tabular}}
  \label{Tab4}
\end{table}

%=======================================================================================
\section{Summary}\label{Sec:Summary}
%=======================================================================================

In summary, the RBHF theory in the full Dirac space has been employed to study the gravitational redshift, moment of inertia, and quadrupole moment of neutron star under the slow-rotation and small-tidal-deformation approximation.
The one-to-one correspondence relation for gravitational redshift and mass is established and used to infer the masses of isolated neutron stars by combining gravitational redshift measurements.
The ratio of the moment of inertia $I$ to $MR^2$ as a function of the compactness $M/R$ is obtained, and is consistent with the universal relations shown by Lattimer et al.~\cite{Lattimer_2005-ApJ629.979} and that from Bayesian posterior probability distributions by Lim et al.~\cite{YeunhwanLim-2019-PhysRevC.100.035802}.
Using $NN$ interactions Bonn A, B, and C, the moment of inertia for $1.338M_\odot$ pulsar PSR J0737-3039A is predicted to be $1.356\times10^{45}$, $1.381\times10^{45}$, and $1.407\times10^{45}\ \mathrm{g~cm^2}$, which are consistent with the constraint translated from the tidal deformability deduced from GW170817 with universal relations among neutron star observables.
The EOSs constructed by the RBHF theory in the full Dirac space, together with those by the projection method and momentum-independence approximation, are compared successfully to universal $I$-Love-$Q$ relations.
By combing the tidal deformability $\Lambda_{1.4M_\odot}$ from GW170817 and the numerical fitting for these universal relations from relativistic \textit{ab initio} EOSs, the moment of inertia of neutron star with 1.4 solar mass is deduced as $I_{1.4M_\odot}=1.22^{+0.40}_{-0.25}\times 10^{45}\mathrm{g\ cm^2}$.

\begin{acknowledgments}

We thank Yeunhwan Lim for providing the probability distribution data for the moment of inertia from the Bayesian analysis and Armen Sedrakian for carefully reading the manuscript.
This work was partly supported by the National Natural Science Foundation of China (NSFC) under Grant No. 12147102, the Fundamental Research Funds for the Central Universities under Grants No. 2020CDJQY-Z003 and No. 2021CDJZYJH-003, and the MOST-RIKEN Joint Project “Ab initio investigation in nuclear physics.”
Part of this work was achieved by using the supercomputer OCTOPUS at the Cybermedia Center, Osaka University under the support of the Research Center for Nuclear Physics of Osaka University.

\end{acknowledgments}

\bibliography{I-Love-Q}
%\end{CJK*}
\end{document}